\documentclass[conference]{IEEEtran}
\IEEEoverridecommandlockouts
\def\BibTeX{{\rm B\kern-.05em{\sc i\kern-.025em b}\kern-.08em
    T\kern-.1667em\lower.7ex\hbox{E}\kern-.125emX}}

\usepackage{indentfirst}
\usepackage{graphicx}
\usepackage{amssymb}
\usepackage{amsfonts}
\usepackage{mathrsfs}
\usepackage{leftidx}
\usepackage{color}
\usepackage{amsmath}
\usepackage{arydshln}
\usepackage{ragged2e}
\usepackage{cite}
\usepackage{enumerate}
\usepackage{longtable}
\usepackage{float}
\usepackage{stfloats}
\usepackage{algpseudocode}
\usepackage{algorithm}
\usepackage{tabularx}
\usepackage{subcaption}
\usepackage{graphicx}

\usepackage[table,usenames,dvipsnames]{xcolor}
\definecolor{aa}{RGB}{175,238,238}
\definecolor{bb}{RGB}{255,255,255}
\usepackage{bm}
\usepackage{makecell}

\begin{document}

\title{SREC: Encrypted Semantic Super-Resolution Enhanced Communication}

\author{
\IEEEauthorblockN{Zhidi Zhang, Rui Meng, Song Gao,  Haixiao Gao, Xiaodong Xu}
\IEEEauthorblockA{ State Key Laboratory of Networking and Switching Technology, BUPT, Beijing, China}

\IEEEauthorblockA{\{2639134068,  buptmengrui, wkd251292, haixiao, xuxiaodong\}@bupt.edu.cn}
\thanks{This work was supported in part by the National Key Research and Development Program of China under Grant 2020YFB1806905; in part by the National Natural Science Foundation of China under Grant 62501066 and under Grant U24B20131; and in part by the Beijing Municipal Natural Science Foundation under Grant L242012. \textit{(Corresponding author: Rui Meng)}}
}

\maketitle

\begin{abstract}
Semantic communication (SemCom), as a typical paradigm of deep integration between artificial intelligence (AI) and communication technology, significantly improves communication efficiency and resource utilization efficiency. However, the security issues of SemCom are becoming increasingly prominent. Semantic features transmitted in plaintext over physical channels are easily intercepted by eavesdroppers. To address this issue, this paper proposes Encrypted Semantic Super-Resolution Enhanced Communication (SREC) to secure SemCom. SREC uses the modulo-256 encryption method to encrypt semantic features, and employs super-resolution reconstruction method to improve the reconstruction quality of images. The simulation results show that in the additive Gaussian white noise (AWGN) channel, when different modulation methods are used, SREC can not only stably guarantee security, but also achieve better transmission performance under low signal-to-noise ratio (SNR) conditions.

\end{abstract}

\begin{IEEEkeywords}
semantic communication, encrypted communication, wireless security.
\end{IEEEkeywords}

\section{Introduction}
As recognized by academia and industry, Artificial Intelligence (AI) will play an indispensable and crucial role in the technological development and evolution process of the sixth generation (6G) mobile communication systems \cite{trichias20246g}.
% On one hand, emerging AI technologies are applied to network planning, maintenance, and optimization, endowing the network with capabilities of self-operation, self-maintenance, and self-repair. On the other hand, networks with native intelligence will be able to provide users with higher-quality AI application services \cite{wang2023on}. 
6G promotes the urgent need for massive data and efficient information analysis \cite{cao2025importance}. However, it is difficult for existing communication systems to meet the needs of emerging 6G applications. \textit{Intellicise (intelligent and concise)} wireless networks, characterized by endogenous intelligence and intrinsic conciseness, have been considered as a promising research direction \cite{wang2024intellicise}.

As a representative technology for intellicise wireless networks, semantic communication (SemCom) leverages AI techniques to extract and transmit information most relevant to the communication objective \cite{lu2025important}. It can not only alleviate the wireless data transmission burden but also improve the efficiency of network control and management \cite{zhang2025intellicise}. 
Currently, joint source-channel coding (JSCC) is considered an important implementation method for SemCom. It breaks the traditional separated source-channel coding approach, directly mapping extracted semantics into feature vectors onto channel symbols, and achieving reliable information transmission by optimizing end-to-end distortion through neural networks \cite{bourtsoulatze2019deep,wu2025lotterycodec}. 

% Currently, the key of SemCom system design is to develop corresponding semantic extraction and joint source-channel coding (JSCC) schemes for different transmission modalities\cite{gao2025cross}.

% However, the security of SemCom is increasingly under threat, specifically as follows: First, current SemCom generally adopts centralized learning to ensure data utilization efficiency. Excessive transparency in information sharing may lead to privacy leakage issues; The data transmitted in SemCom essentially contains contextual information, and related sensitive information  may be inadvertently leaked \cite{do2025international}. Second, Attackers can intercept transmitted signals and use generative AI models to infer semantic information. In addition, attackers can not only attack semantic transmission information but also attack machine learning models used for semantic extraction, reducing the practicality of the models \cite{yang2024secure}.
However, the security of SemCom is increasingly being threatened, as follows: firstly, semantic features, as the data carrier of SemCom, directly carry the key information of users, and information about tasks, senders, and even the data itself may be inadvertently leaked\cite{do2025security}. Secondly, the openness of wireless channels poses significant security risks to the semantic features of plaintext transmission, making it susceptible to eavesdropping attacks and sensitive information leakage\cite{qin2023securing}. In addition, in the JSCC scenario, it becomes difficult to apply traditional encryption schemes that rely on source channel separation coding, and it is necessary to use encryption methods compatible with JSCC while ensuring its performance\cite{rong2025semantic,wu2025actions}.

In response to the above security threats, researchers employ encryption techniques to enhance the confidentiality and integrity of SemCom, such as classic cryptography algorithms\cite{tung2023deep}, homomorphic encryption\cite{meng2025secure}, and quantum cryptography\cite{kaewpuang2024cooperative}. Motivated by physical-layer security, Zhao et al. introduce physical-layer keys for securing SemCom\cite{zhao2022semkey}. Also, some researchers consider covert communications\cite{xu2024covert} and steganography techniques\cite{tang2024secure} to defend against semantic eavesdroppers. Although the above schemes have proposed possible solutions to the existing problems, there are still challenges in implementing lightweight and secure SemCom.

Against the above background, we introduce super-resolution to enhance the transmission performance for encrypted SemCom systems. 
Super-resolution aims to recover high-resolution images with more details and higher clarity from low-resolution images\cite{yang2020a}. 
Traditional super-resolution methods mainly include interpolation and sparse coding. Currently, deep learning-based super-resolution reconstruction has attracted much attention. For example, Wu et al. \cite{wu2024seesr} propose a semantics-aware method to enable the reconstructed image to reproduce more realistic image details and better preserve image semantics. 
Therefore, we propose an Encrypted Semantic Super-Resolution Enhanced Communication (SREC) scheme. The main contributions are summarized as follows.
\begin{itemize}
    \item We propose SREC, a secure SemCom method that integrates cryptographic encryption methods to protect image semantic transmission from eavesdropping.
    \item To address the bit error impact that may be caused by the encryption and decryption processes, we introduce a super-resolution reconstruction module, which improves the image reconstruction effect, especially under low Signal-to-Noise Ratio (SNR) environments.
    \item We conducted experiments on the Urban100\cite{huang2015single} dataset using Peak Signal to Noise Ratio (PSNR) as the evaluation metric. And we use this to verify the effectiveness of our proposed SREC.

\end{itemize}

\section{The Proposed SREC Scheme}
This section first outlines the overall architecture of the proposed SREC scheme, and then elaborates on its key modules, including the modulo-256 encryption and super-resolution-based semantic enhanced modules. 

\subsection{Overall Architecture of SREC Scheme}
\begin{figure*}[h]
     \centering
     \includegraphics[scale=0.5]{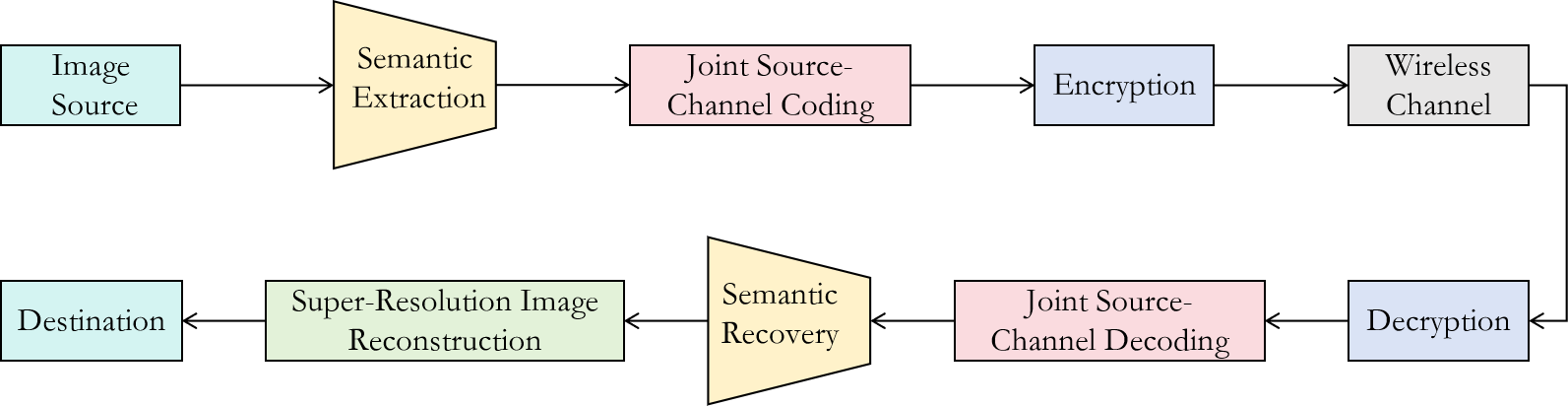} 
     \caption{The architecture of the proposed SREC, where semantic extraction, JSCC encoding, and encryption on the input image are performed at the transmitter, and decryption, JSCC decoding, semantic recovery, and super-resolution reconstruction are performance ar the receiver.}
     \label{The overall architecture of SREC}
\end{figure*}

As shown in Figure \ref{The overall architecture of SREC}, the proposed SREC scheme includes the following modules.

% \begin{figure*}[htbp]
%      \centering
%      \includegraphics[scale=0.3]{total.png} 
%      \caption{Overall architecture of SREC: Input image undergoes semantic extraction, JSCC encoding, and encryption before transmission; receiver performs decryption, JSCC decoding, semantic recovery, and final super-resolution reconstruction.}
%      \label{fig:srec_overall_architecture}
% \end{figure*}

\begin{itemize}
\item 
\textit{Semantic Extraction:}
The input image $x$ undergoes neural network-based semantic extraction to form a semantic vector $y$:
\begin{equation}\label{Ga}
    y = G_a(x),
\end{equation}
where $G_a(\cdot)$ denotes the semantic extraction operation.
\item 
\textit{Joint Source-Channel Coding:}
The extracted semantic vector $y$ undergoes joint source-channel coding to form a vector $s$:
\begin{equation}\label{Fe}
    s = F_e(y),
\end{equation}
where $F_e(\cdot)$ denotes the joint source-channel coding operation.
\item 
\textit{Encryption:}
The encoded vector $s$ is encrypted to form an encrypted vector $s_{\text{enc}}$:
\begin{equation}\label{Enc}
    s_{\text{enc}} = Enc(s),
\end{equation}
where $Enc(\cdot)$ denotes the encryption operation.
\item 
\textit{Physical Channel:}
The physical channel is modeled as Additive White Gaussian Noise (AWGN), and its expression is as $n \sim \mathcal{C} \mathcal{N}(0, \sigma)$. Here, \(\sigma\) denotes the noise power. Therefore, the encrypted signal received at the receiver is expressed as:
\begin{equation}\label{h}
    s_{\text{trans}}=h*s_{\text{enc}}+n,
\end{equation}
where $h$ represents the coefficient of the physical channel between the transmitter and the receiver.
\item 
\textit{Decryption:}
The encrypted vector received from the physical channel is decrypted to form a decrypted vector $s_{\text{dec}}$:
\begin{equation}\label{Dec}
    s_{\text{dec}} = Dec(s_{\text{trans}}),
\end{equation}
where $Dec(\cdot)$ denotes the decryption operation.
\item 
\textit{Joint Source-Channel Decoding:}
The decrypted vector is decoded to obtain a reconstructed semantic vector $\hat{y}$:
\begin{equation}\label{Fd}
    \hat{y} = F_d(s_{\text{dec}}),
\end{equation}
where $F_d(\cdot)$ denotes the joint source-channel decoding operation. The decoder structure is similar to that of the encoder, and the decoder can be regarded as the inverse structure of the encoder.
\item 
\textit{Semantic Recovery:}
Semantic recovery is performed on the reconstructed semantic vector to obtain a recovered input image $\hat{x}$:
\begin{equation}\label{Gs}
    \hat{x} = G_s(\hat{y}),
\end{equation}
where $G_s(\cdot)$ denotes the semantic recovery operation. The structure of semantic recovery is similar to semantic extraction, and semantic recovery can be regarded as the inverse process of semantic extraction.
\item 
\textit{Super-Resolution Reconstruction:}
Super-resolution reconstruction is performed on the decoded data to obtain the final output reconstructed image $\hat{x}_{\text{sr}}$:
\begin{equation}\label{rdn}
    \hat{x}_{\text{sr}} = RDN(\hat{x}),
\end{equation}
where $RDN(\cdot)$ denotes the super-resolution reconstruction operation. The process of SREC is provided in Algorithm \ref{algorithm}.
\end{itemize}

\begin{algorithm}
    \caption{Encrypted Semantic Super-Resolution Enhanced Communication}
    \label{algorithm}
    
    \begin{algorithmic}[1]
        \Require Input image $x$, Channel coefficient $h$, Noise power $\sigma$, Encryption key, \textbf{KEY} (pre-shared)
 
        \State Extract semantic vector from $x$ by (\ref{Ga})
        
        \State JSCC encode semantic vector for channel transmission by (\ref{Fe})
        
        \State Encrypt JSCC encoded vector with \textbf{KEY} by (\ref{Enc})
        
        \State Transmit encrypted vectors through physical channels by (\ref{h})
        
        \State Decrypt the received encrypted vector with \textbf{KEY} by (\ref{Dec})
        
        \State JSCC decode decrypted vector to reconstruct semantic vector by (\ref{Fd})
        
        \State Perform semantic recovery on the decoded vector by (\ref{Gs})
        
        \State Perform super-resolution reconstruction on the reconstructed image by (\ref{rdn})
        
        \State \Return $\hat{x}_{\text{sr}}$
    \end{algorithmic}
\end{algorithm}

\subsection{Modulo-256-based Encryption Module}
Modulo-256 encryption is an encryption idea based on modulo-256 operations, which involves performing a certain mathematical operation between each byte of the plaintext and the corresponding byte of the key, then taking modulo 256 of the result\cite{wang2023a}. The proposed SREC performs modulo-256 encryption on each channel of the normalized feature tensor $s$ obtained after joint source-channel coding of the extracted semantic features. The key used for encryption is a pseudorandom tensor equal in size to the feature tensor. Each element in the key is a torch.uint8 type element in the range from 0 to 255, denoted as:
\begin{equation}
    \textbf{KEY} = \{KEY(i)|i = 1,2,...,length, KEY(i)\in[0,255]\},
\end{equation}
where length is the number of elements in \textbf{KEY}. The encryption of the vector s using the \textbf{KEY} is performed as follows:
\begin{equation}\label{encrypt}
    s_{enc}(i) = (s(i) + KEY(i) \mod 256).
\end{equation}
Since there currently exists no algorithm capable of probabilistically distinguishing pseudorandom sequences from random sequences in polynomial time, adversaries cannot distinguish $s_{enc}$ at any time; thus, the encryption method in Equation \ref{encrypt} is semantically secure \cite{zhang2012scalable}.

After encryption,$s_{enc}$ is transmitted over the public channel, while \textbf{KEY} is transmitted to the receiver via the secure channel.

\subsection{Super-Resolution Based Semantic Enhanced Module}
\begin{figure}[]
     \centering
     \includegraphics[scale=0.5]{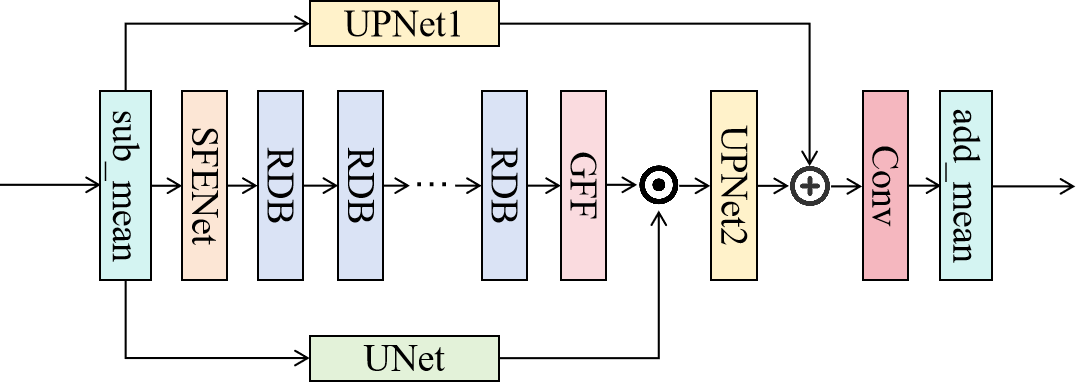} 
     \caption{The architecture of the super-resolution based semantic enhanced module.}
     \label{RDN}
\end{figure}

This paper adopts a super-resolution reconstruction network based on Residual Dense Network (RDN) \cite{zhang2018residual} to improve image reconstruction quality and enhance semantics, and the workflow of this network is shown in Figure \ref{RDN}.

First, preprocessing is performed on the input image $I_{LR}$ to achieve color normalization:
\begin{equation}
    x_0 = sub\_mean(I_{LR}).
\end{equation}

The UNet network is used to process $x_0$, obtaining the channel attention weight map of $x_0$, denoted as $weight$, which is used for subsequent implementation of channel attention:
\begin{equation}
    weight = UNet(x_0),
\end{equation}
where $UNet(\cdot)$ employs four downsampling blocks and four upsampling blocks, preserving image details by connecting corresponding layers. 

$H_{up1}(\cdot)$ is used to process $x_0$, obtaining the initial feature $f_1$ of $x_0$, which is used for subsequent residual connections:
\begin{equation}
    f_1 = H_{up1}(x_0),
\end{equation}
where $H_{up1}(\cdot)$ consists of one convolutional layer and one pixel shuffle. 

$H_{sfe}(\cdot)$ is used to process $x_0$, obtaining the shallow feature $F_0$ of $x_0$, which provides basic feature representations for subsequent deep feature extraction:
\begin{equation}
    F_0 = H_{sfe}(x_0),
\end{equation}
where $H_{sfe}(\cdot)$ is composed of two convolutional layers.

The core of RDB is to maximize the transmission of information between different layers within the network. The input of each convolutional layer in an RDB includes the initial output of the RDB and the outputs of all previous convolutional layers within the same RDB. These feature maps are aggregated through channel concatenation to form the output of the next layer, thereby realizing continuous transmission and reuse of features. Shallow feature $F_0$ is processed using RDBs. For each RDB where $d = 1, 2, \ldots, D$:
\begin{equation}
    F_d = H_{RDB_d}(F_{d-1}).
\end{equation}

The internal operation of each RDB can be expressed as:
\begin{align}
    F_{d,c} &= \sigma(W_{d,c} \ast [F_{d-1}, F_{d,1}, \ldots, F_{d,c-1}]), \\
    F_{d,\text{LF}} &= W_{d,\text{LFF}} \ast [F_{d-1}, F_{d,1}, \ldots, F_{d,C}], \\
    F_d &= F_{d-1} + F_{d,\text{LF}},
\end{align}
where $\sigma$ is the ReLU activation function, $W_{d,c}$ is the weight of the $c$-th convolutional layer in the $d$-th RDB, and $W_{d,\text{LFF}}$ is the $1 \times 1$ convolutional weight for local feature fusion. RDN contains multiple RDBs, each extracting features with different depths and receptive fields: shallow RDBs capture local details and textures of the image, while deep RDBs capture more global and abstract feature information.

After extracting local features through a series of RDBs, the network further proposes dense feature fusion (DFF) to explore multi-level features from a global perspective. DFF mainly consists of two parts: global feature fusion (GFF) and global residual learning (GRL). GFF fuses RDB features through convolution:
\begin{equation}
    F_{gf} = H_{GFF}([F_1, F_2, \ldots, F_D]),
\end{equation}
where $H_{GFF}(\cdot)$ can generate richer and more accurate feature representations than a single RDB by integrating features extracted by all RDBs. It effectively fuses local features output by all RDBs, avoiding the limitation of the network only using the deepest features, and ensuring that both shallow texture details and deep contextual information can contribute to the final reconstruction process.

The feature map $F_{gf}$ after GFF is multiplied element-wise with the weight map $weight$:
\begin{equation}
    F_{mod} = F_{gf} \times weight,
\end{equation}
where element-wise multiplication of $F_{gf}$ and $weight$ enables the spatial attention mechanism for features, enhancing feature responses in important regions and suppressing noise in unimportant regions, allowing the network to adaptively process different regions of the image.

The feature-modulated feature map $F_{mod}$ is upsampled:
\begin{align}
    F_{up} &= H_{up2}(F_{mod}),
\end{align}
where similar to $H_{up1}(\cdot)$, $H_{up2}(\cdot)$ consists of one convolutional layer and one pixel shuffle. Upsampling can map low-resolution features to the target high resolution.

GRL adds the extracted initial feature $f_1$ and the deep feature $F{up}$ processed by the network:
\begin{align}
    F_{res} &= F_{up} + f_1.
\end{align}
This ensures that the basic information and structure of the original image are not completely lost during deep processing, allowing the entire network to focus on learning the differences between shallow features and the target output, thereby simplifying the learning task.

A convolutional layer is used for final feature refinement:
\begin{align}
    F_{final} &= W_{final} \ast F_{res},
\end{align}
where $W_{final}$ is the weight of the final convolutional layer.

Finally, the previously subtracted mean is added to the obtained image to get the final super-resolution image $I_{HR}$:
\begin{align}
    I_{SR} &= {add\_mean}(F_{final}).
\end{align}

After super-resolution reconstruction processing, the reconstructed image $I_{SR}$ not only has richer detail information compared to the original input image $I_{LR}$, but also contains clearer semantic information.

\section{Experimental Results And Analysis}
% This section employs on various parameter configurations adopted in the simulation, then analyzes the specific simulation results to verify and evaluate the performance of the proposed encrypted transmission model for SemCom.
This section presents the training and testing datasets, parameter settings, and analysis of experimental results. Throught simulation analysis, we have verified the performance of the proposed SREC.

\subsection{Simulation Parameters}
\subsubsection{Datasets}
We select the DIV2K\cite{agustsson2017ntire} dataset as the training set and validation set respectively, and at the same time select 25 images from the Urban100 dataset to form the test set. To ensure the consistency of experimental conditions, all experimental images are uniformly cropped to a size of 1024 × 512 pixels.

\subsubsection{Parameter setting}
The model proposed in this paper uses Python as the programming language, with version 3.11.4; the deep learning framework employed is PyTorch, with version 2.5.1; and the graphics card used is the NVIDIA RTX 6000 Ada Generation. In addition, simulation selects nonlinear transform source-channel coding (NTSCC)\cite{dai2022nonlinear} as the semantic extraction and JSCC network. The network is trained with the condition of SNR of 10dB.
% The image dataset for training and validation is DIV2K\cite{agustsson2017ntire}. In addition, 25 images selected from Urban100 are used as the test set. Images in the training set, validation set, and test set are all cropped to a size of 512×1024. The training hyperparameters are shown in \ref{Hyperparameters}.
\begin{table}[]
    \centering
    \caption{Hyperparameters}
    \begin{tabular}{|>{\centering\arraybackslash}m{3cm}|>{\centering\arraybackslash}m{3cm}|}
    \hline
    \textbf{Hyperparameter} & \textbf{Value}        \\
    \hline
    epoch          & 50           \\
    \hline
    test every    & 1000         \\
    \hline
    batch size    & 8            \\
    \hline
    learning rate  & 5e-5         \\
    \hline
    learning rate decay factor for step decay $\gamma$       & 0.5          \\
    \hline
    optimizer      & Adam         \\
    \hline
    ADAM beta $\beta$        & (0.9, 0.999) \\
    \hline
    ADAM epsilon for numerical stability $\varepsilon$  & 1e-8         \\
    \hline
    weight decay  & 0            \\
    \hline
    gradient clipping threshold gclip          & 0            \\
    \hline
    loss           & MSE          \\
    \hline
\end{tabular}
    \label{Hyperparameters}
\end{table}

\subsection{Simulation Results}
\subsubsection{Performance Under Different SNRs}
\begin{figure}[]
    \centering 
    
    \begin{subfigure}{\linewidth}
        \centering
        \includegraphics[width=0.7\linewidth]{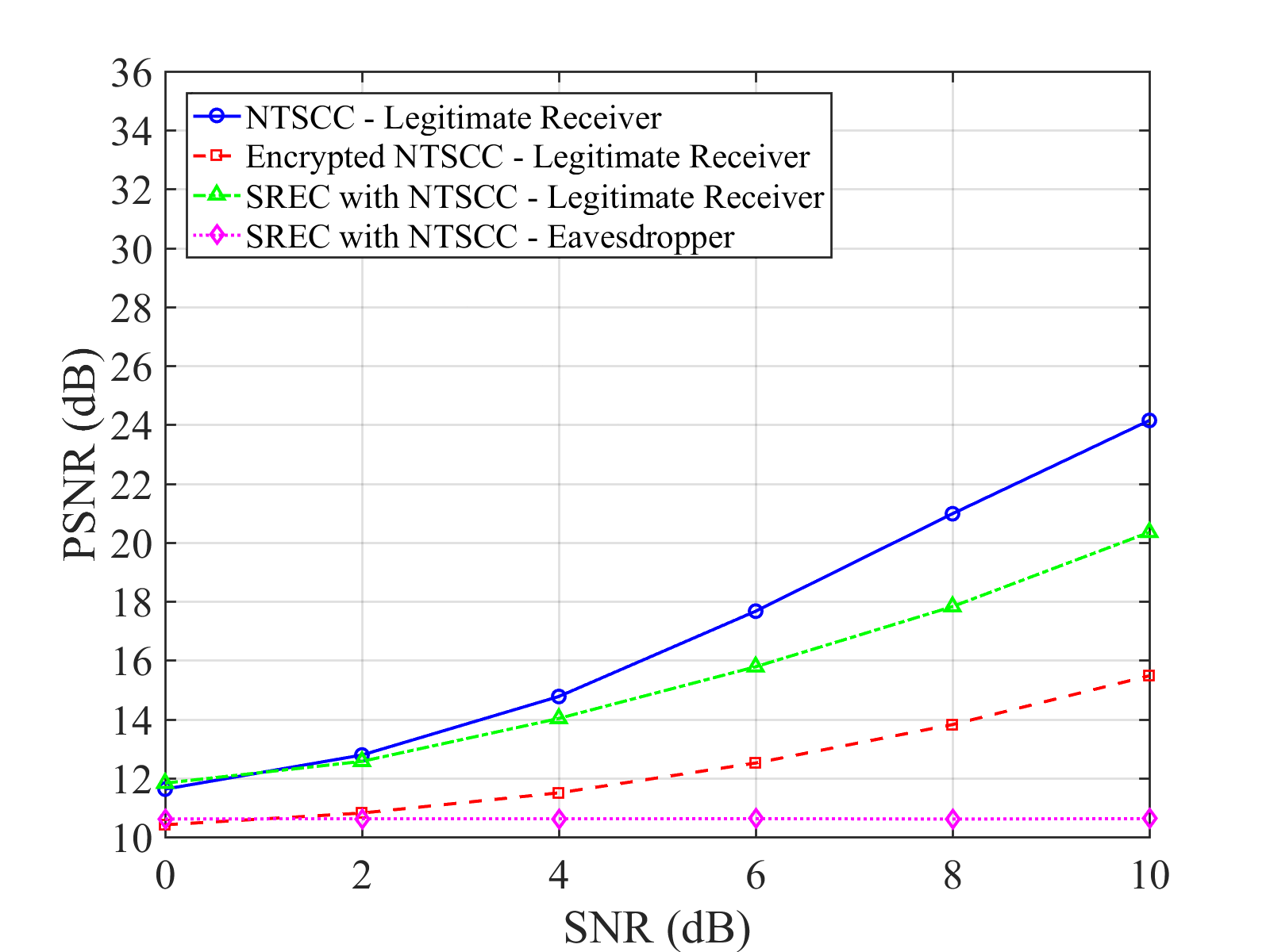} 
        \caption{PSNR-SNR curve with $\eta=0.2$ under 16QAM modulation scheme}
        \label{snr_16qam}
    \end{subfigure}
    
    \begin{subfigure}{\linewidth}
        \centering
        \includegraphics[width=0.7\linewidth]{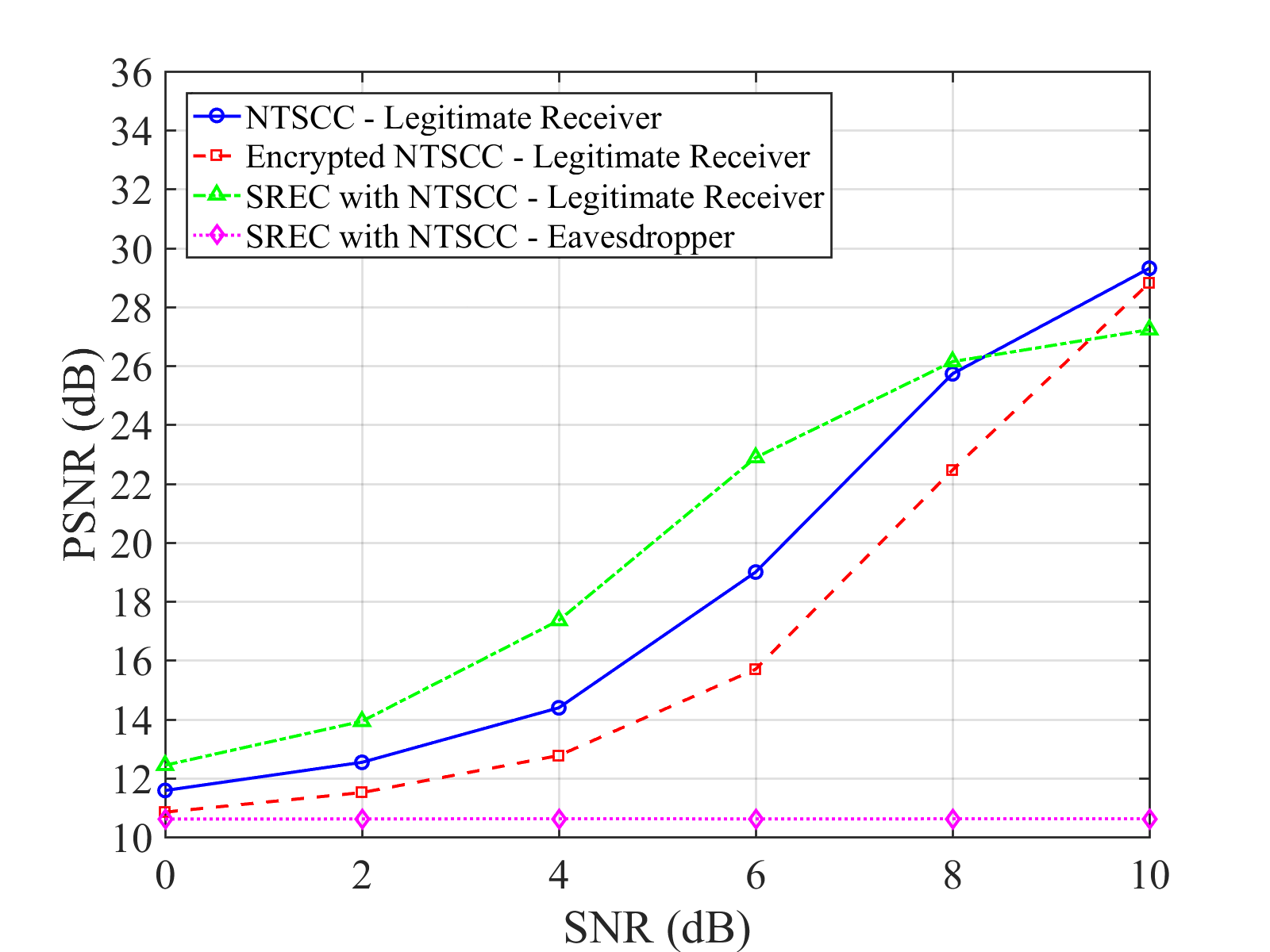}
        \caption{PSNR-SNR curve with $\eta=0.2$ under QPSK modulation scheme}
        \label{snr_qpsk}
    \end{subfigure}
    
    \begin{subfigure}{\linewidth}
        \centering
        \includegraphics[width=0.7\linewidth]{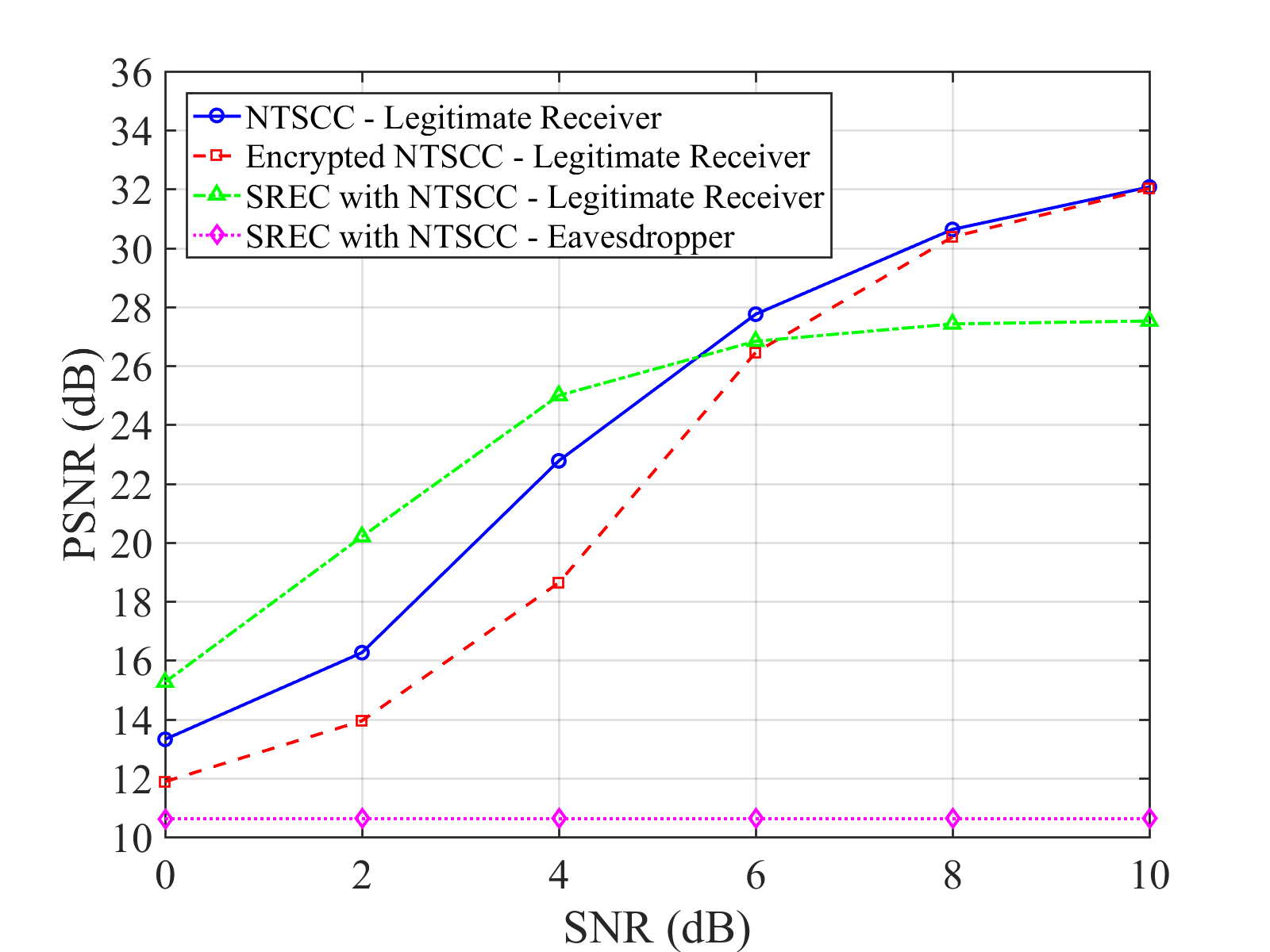}
        \caption{PSNR-SNR curve with $\eta=0.2$ under BPSK modulation scheme}
        \label{snr_bpsk}
    \end{subfigure}

    \caption{PSNR-SNR curves with $\eta=0.2$ and different modulation schemes}
    \label{fig:psnr-snr-curves}
\end{figure}

\begin{figure}[]
    \centering 
    
    \begin{subfigure}{\linewidth}
        \centering
        \includegraphics[width=0.8\linewidth]{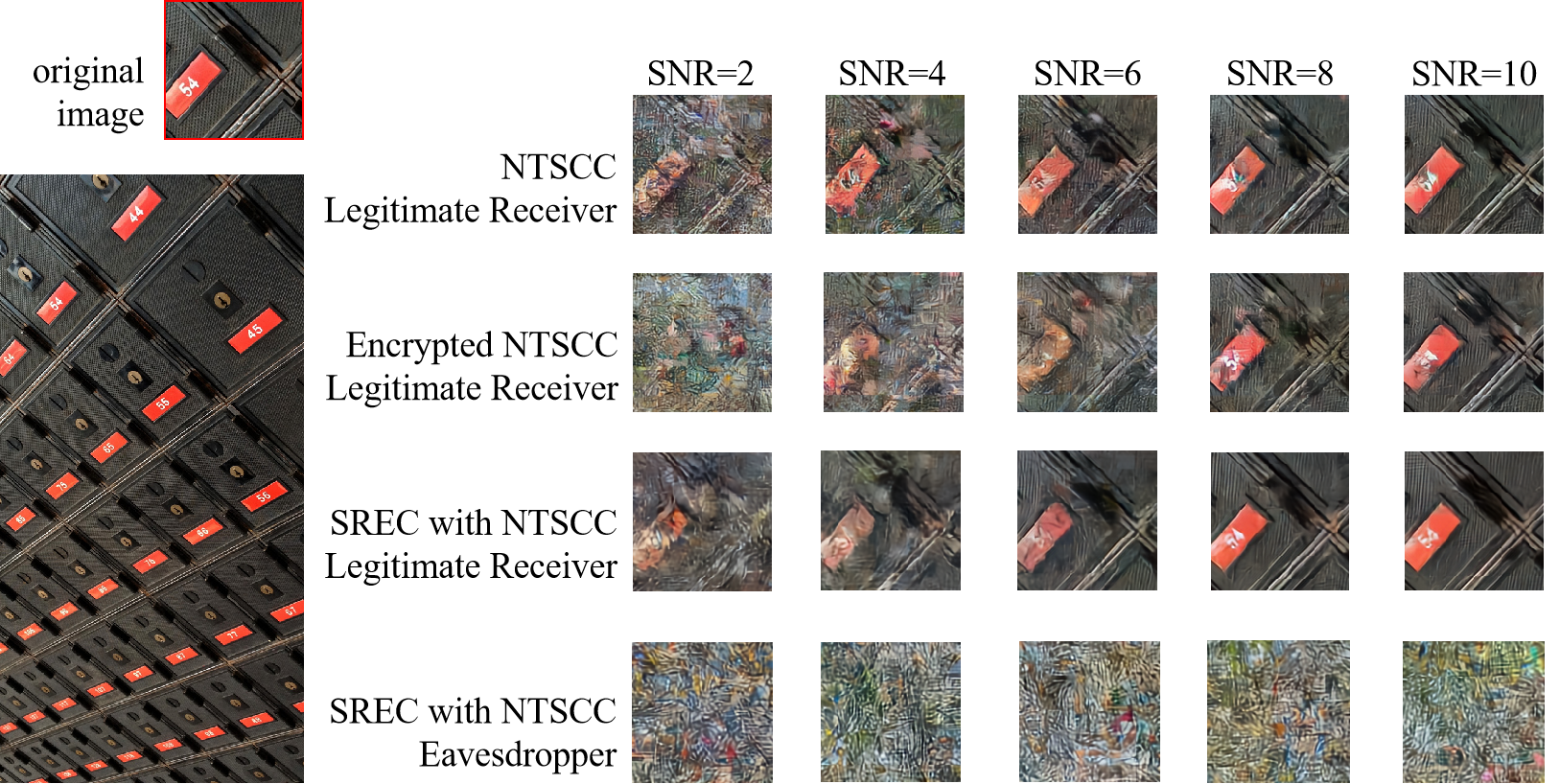} 
        \caption{Visualization images with $\eta=0.2$ and 16QAM modulation scheme}
        \label{fig:16qam}
    \end{subfigure}
    
    \begin{subfigure}{\linewidth}
        \centering
        \includegraphics[width=0.8\linewidth]{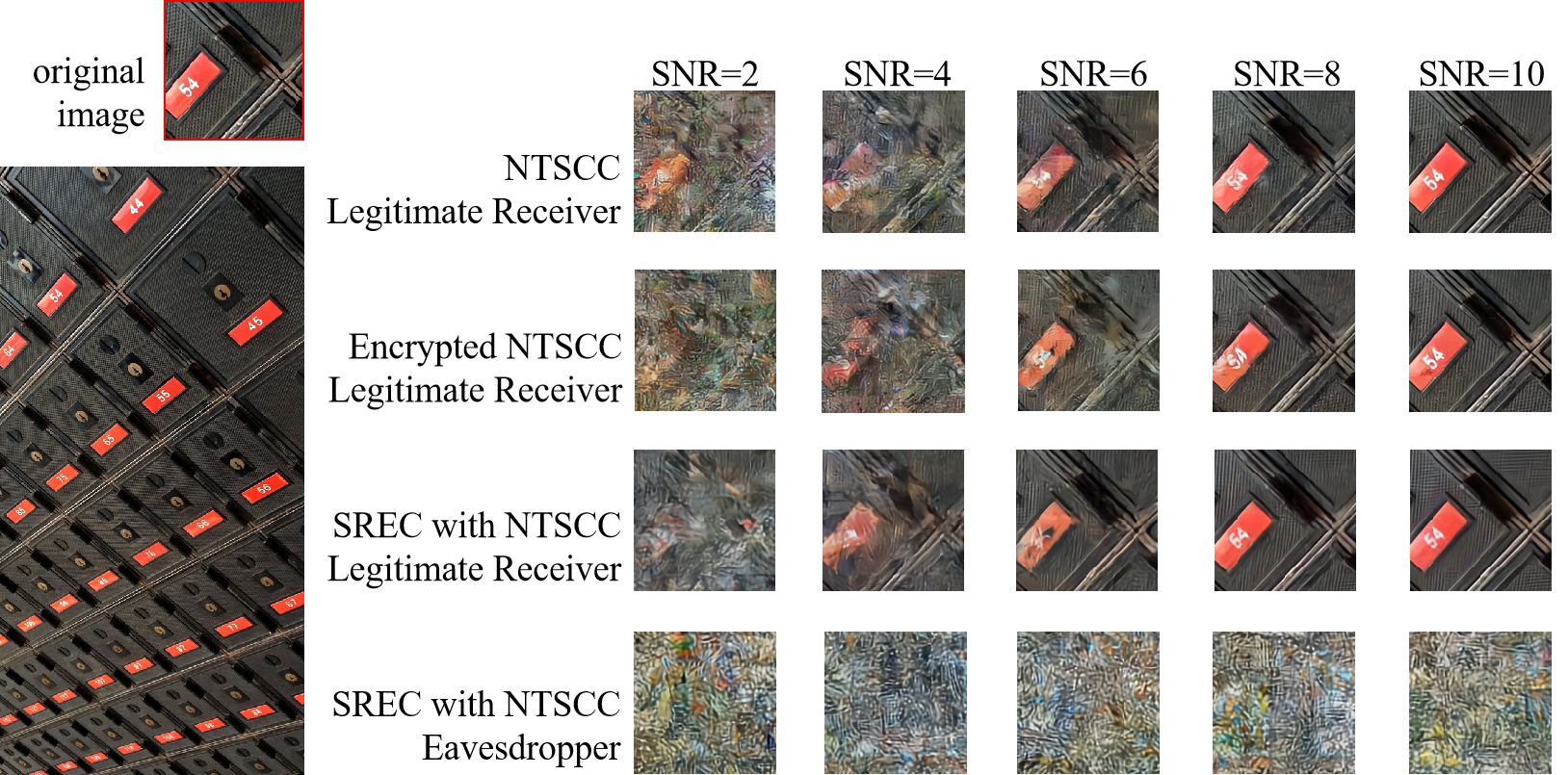}
        \caption{Visualization images with $\eta=0.2$ and QPSK modulation scheme}
        \label{fig:qpsk}
    \end{subfigure}
    
    \begin{subfigure}{\linewidth}
        \centering
        \includegraphics[width=0.8\linewidth]{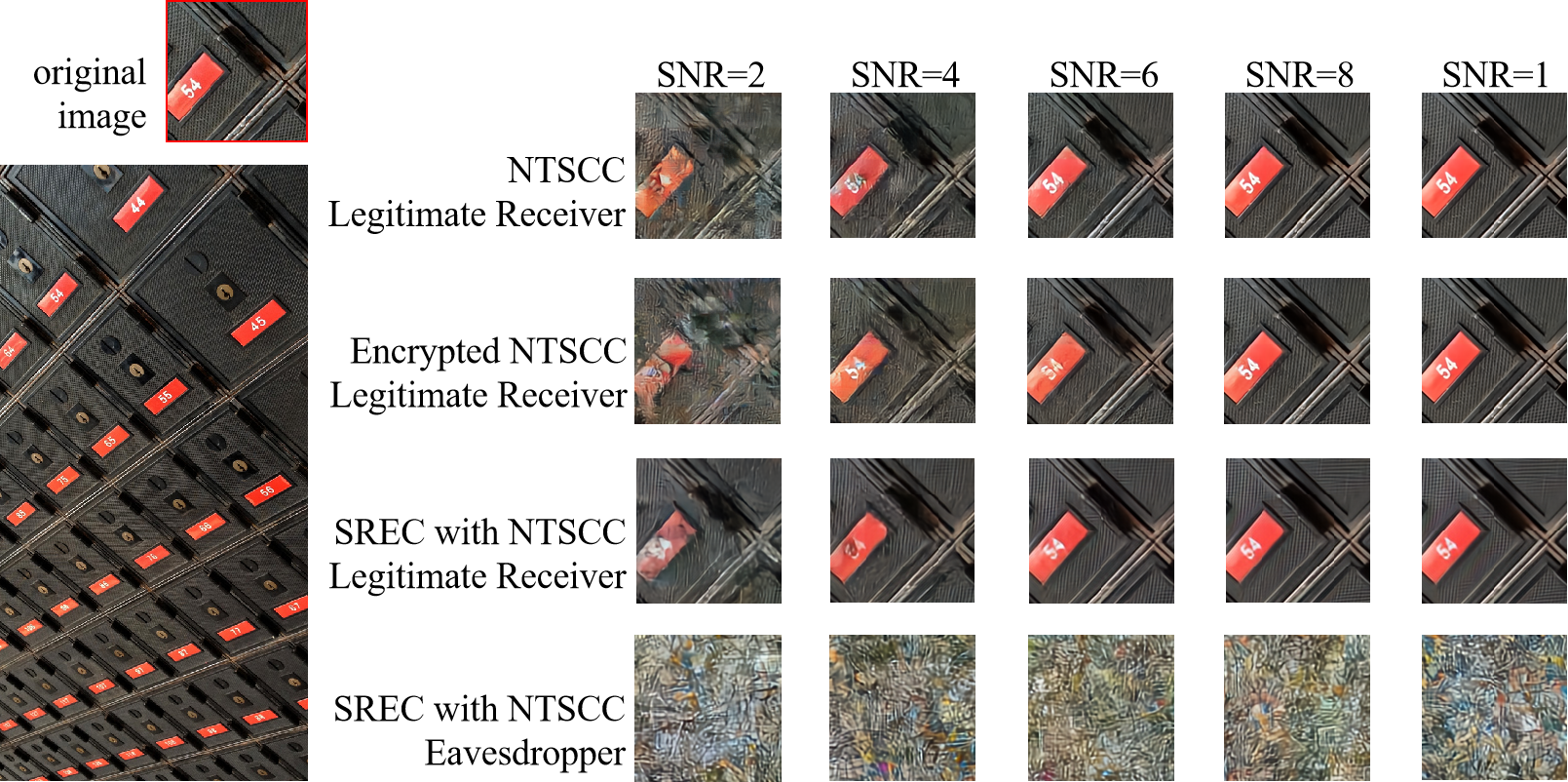}
        \caption{Visualization images with $\eta=0.2$ and BPSK modulation scheme}
        \label{fig:bpsk}
    \end{subfigure}

    \caption{Visualization images with $\eta=0.2$ and  different modulation schemes}
    \label{fig:psnr-snr-visual}
\end{figure}

Under different modulation schemes (16QAM, QPSK, BPSK), simulations were conducted over an AWGN channel with $\eta$ = 0.2 to obtain the relation between PSNR and channel SNR for four schemes, as well as visualized image reconstruction results as shown in Figure \ref{fig:psnr-snr-curves} and \ref{fig:psnr-snr-visual}.

In terms of modulation schemes, different modulation significantly affect image reconstruction quality under the same SNR. In general, low-order modulation has strong noise resistance, enabling reliable transmission of a small amount of information even at low SNR, resulting in higher reconstructed image quality; high-order modulation has a higher bit error rate under the same SNR condition and requires a higher SNR to achieve the same reliability as low-order modulation methods. Figure \ref{snr_16qam} shows that 16QAM exhibits poor image quality at low SNR; as SNR increases, the reconstruction quality of 16QAM improves slowly but remains significantly lower than that of QPSK and BPSK. Figure \ref{snr_bpsk} shows that BPSK has the strongest noise resistance: when SNR increases from 0 to 4 dB, the slope of the curve is significantly greater than that of QPSK and 16QAM. This indicates that even with high channel noise, recognizable image contours can be restored, and a small increase in SNR can lead to a noticeable improvement in image quality. However, when SNR exceeds 7 dB, BPSK almost reaches saturation, and the improvement in image quality slows down. Figure \ref{snr_qpsk} shows that QPSK has a transmission rate between 16QAM and BPSK under the same channel bandwidth; its performance at medium and low SNR is better than 16QAM but worse than BPSK, and QPSK requires a higher SNR to achieve the same reconstruction effect as BPSK.

In terms of model structure, the introduction of encryption and decryption modules will amplify the impact of bit errors under low SNR conditions, reducing image quality; under high SNR conditions, the encryption and decryption processes have almost no impact on reconstruction results. From the perspective of eavesdropping attacks, assuming an eavesdropper holds all model parameters but lacks the key, as shown in the baseline scheme-eavesdropper curve. Without the key for decryption, even if encrypted information is successfully transmitted, the image content cannot be restored.

The super-resolution module is introduced to improve the resolution and visual quality of received images, especially under transmission constraints. It can be clearly seen from the curves that under low SNR conditions, the super-resolution module significantly improves the PSNR of images—particularly for QPSK with SNR below 8 dB and BPSK with SNR below 5 dB, the model with the super-resolution module even achieves a better PSNR than the model without encryption and decryption modules. However, as SNR increases, the image quality gain from the super-resolution module decreases, and may even be lower than that without the super-resolution module. This indicates that although the super-resolution module can reconstruct details, when channel conditions are sufficiently good, the image details "guessed" through inference by the super-resolution module are ultimately less accurate than the original details transmitted directly.

A specific numerical analysis of the three modulation schemes at a representative low SNR of 4 dB shows that the PSNR of reconstructed images using 16QAM-modulated SREC is similar to that of NTSCC, and 2.5 dB higher than that of NTSCC encrypted with the same encryption method; the PSNR of reconstructed images using QPSK-modulated SREC is 3.0 dB higher than that of NTSCC and 4.6 dB higher than that of NTSCC encrypted with the same encryption method; the PSNR of reconstructed images using BPSK-modulated SREC is 2.2 dB higher than that of NTSCC and 6.4 dB higher than that of NTSCC encrypted with the same encryption method. These data demonstrate that under low signal-to-noise ratio conditions, SREC can significantly improve the quality of reconstructed images, and even achieve better reconstructed images than unencrypted transmission.

\subsubsection{Performance Under Different Channel Scaling Factors}
\begin{figure}[]
    \centering 
    
    \begin{subfigure}{\linewidth}
        \centering
        \includegraphics[width=0.7\linewidth]{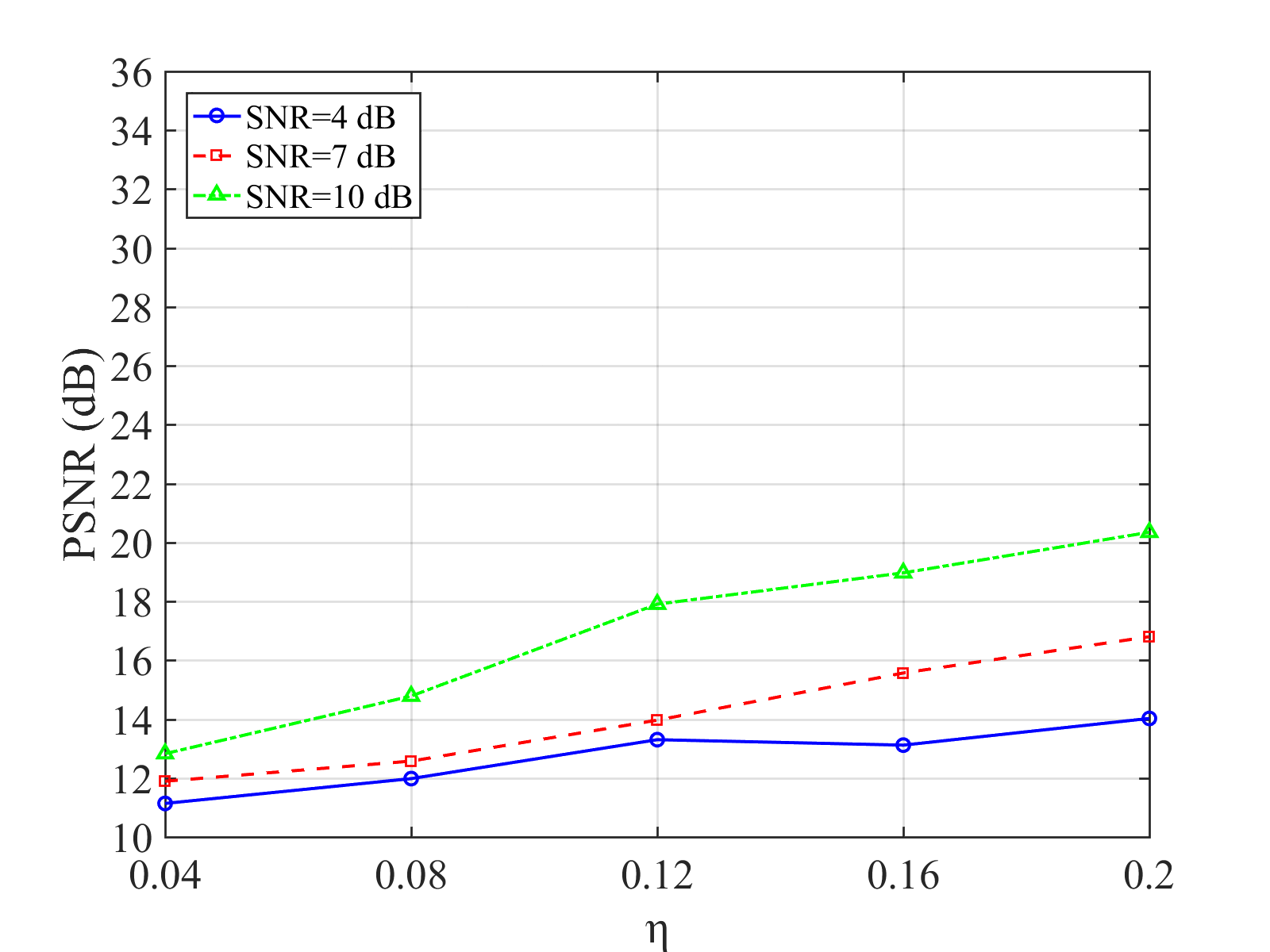} 
        \caption{PSNR-$\eta$ curve with 16QAM scheme}
        \label{eta-16qam}
    \end{subfigure}
    
    \begin{subfigure}{\linewidth}
        \centering
        \includegraphics[width=0.7\linewidth]{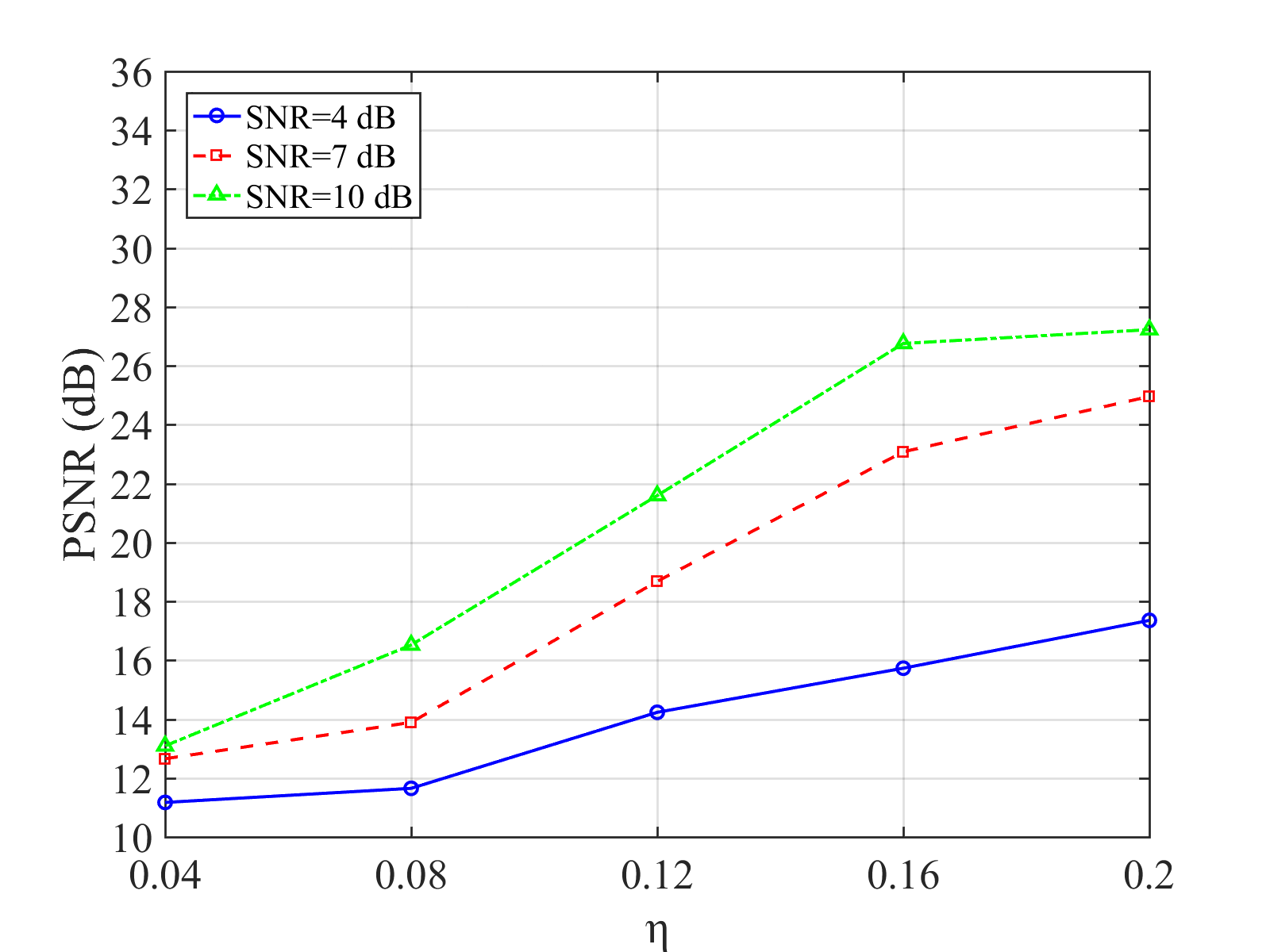}
        \caption{PSNR-$\eta$ curve with QPSK scheme}
        \label{eta-qpsk}
    \end{subfigure}
    
    \begin{subfigure}{\linewidth}
        \centering
        \includegraphics[width=0.7\linewidth]{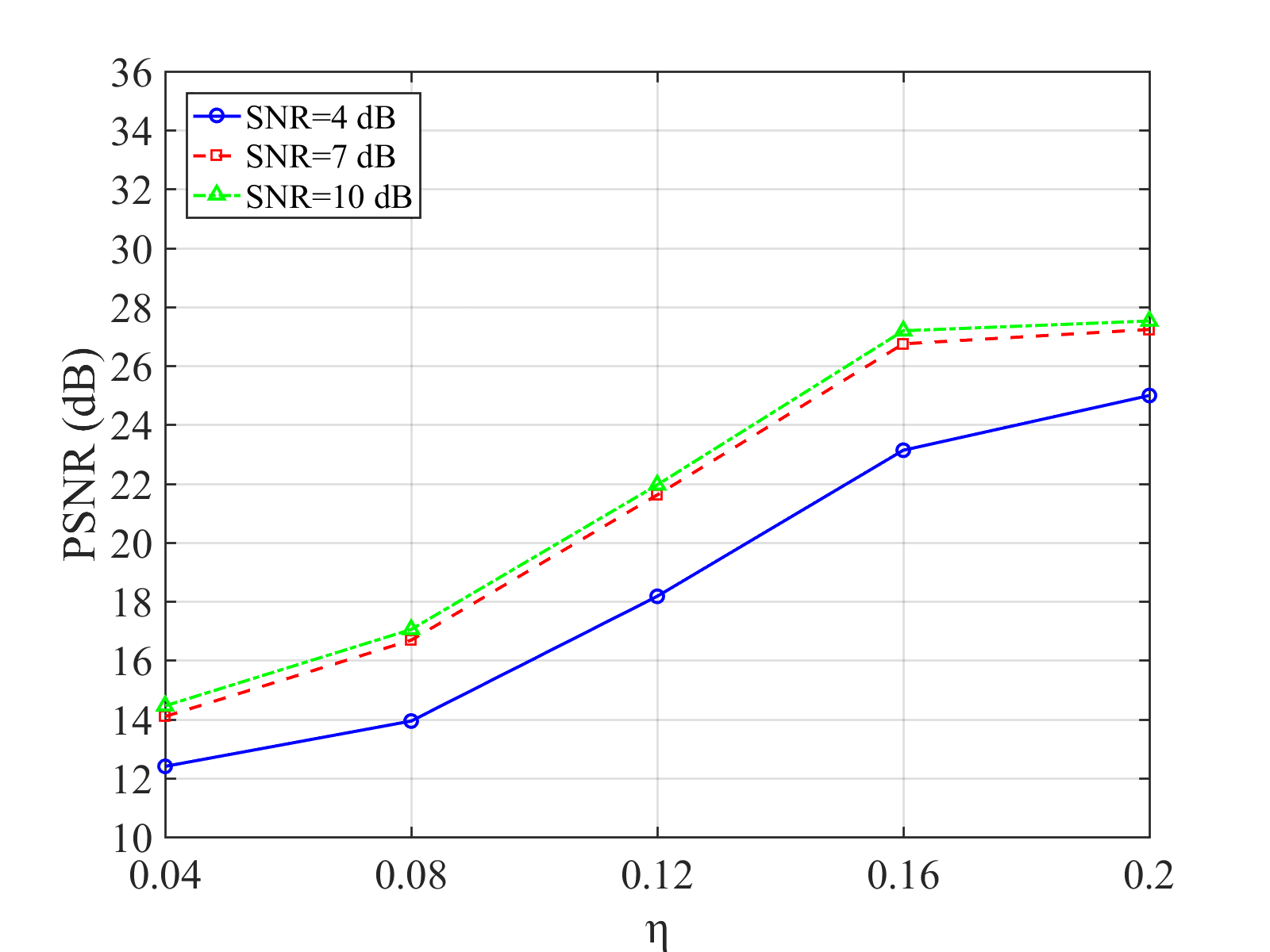}
        \caption{PSNR-$\eta$ curve with BPSK scheme}
        \label{eta-bpsk}
    \end{subfigure}

    \caption{PSNR-$\eta$ curves with different modulation schemes}
    \label{fig:psnr-eta-curves}
\end{figure}

\begin{figure}[]
    \centering 
    
    \begin{subfigure}{\linewidth}
        \centering
        \includegraphics[width=0.8\linewidth]{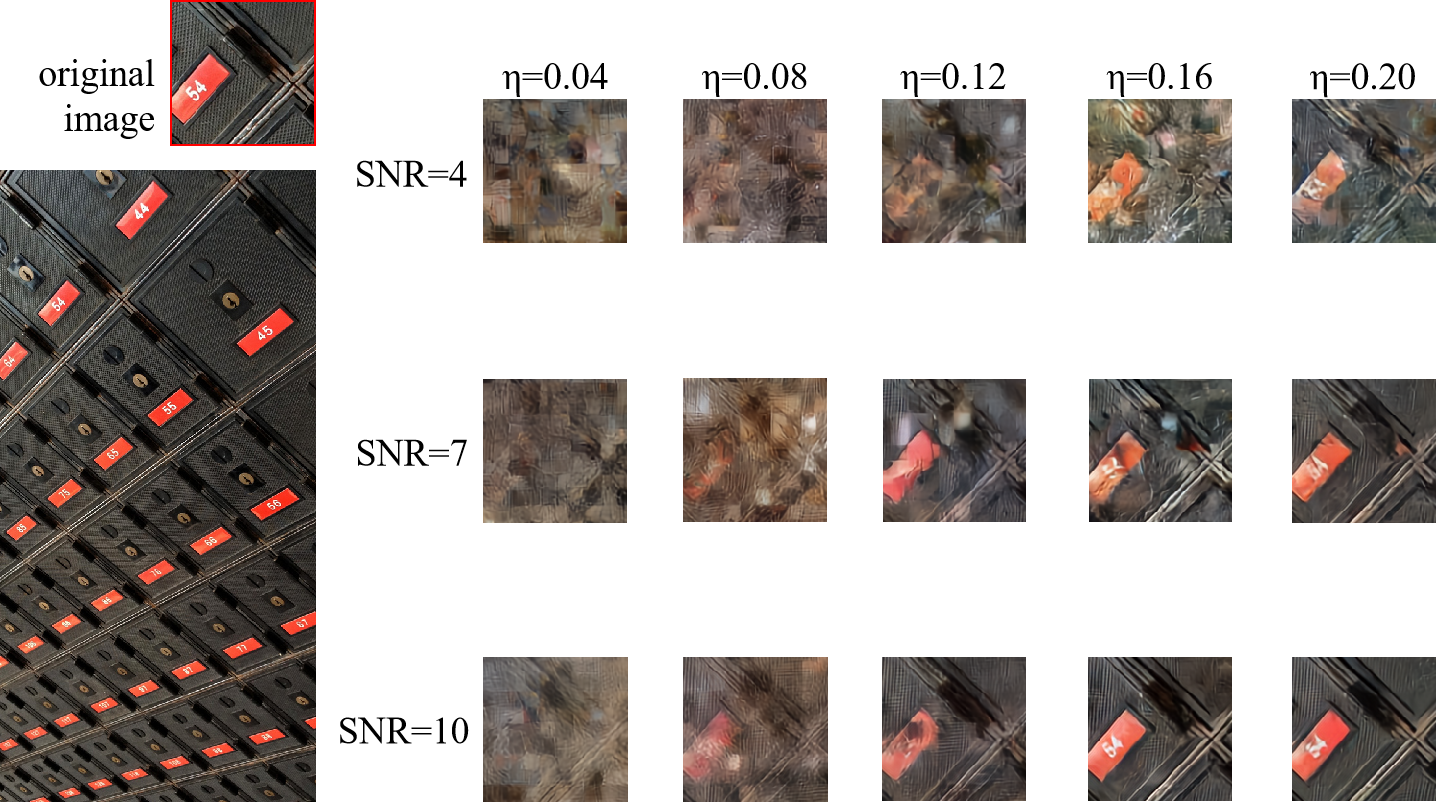} 
        \caption{Visualization images with 16QAM modulation scheme}
        \label{fig:16qam}
    \end{subfigure}
    
    \begin{subfigure}{\linewidth}
        \centering
        \includegraphics[width=0.8\linewidth]{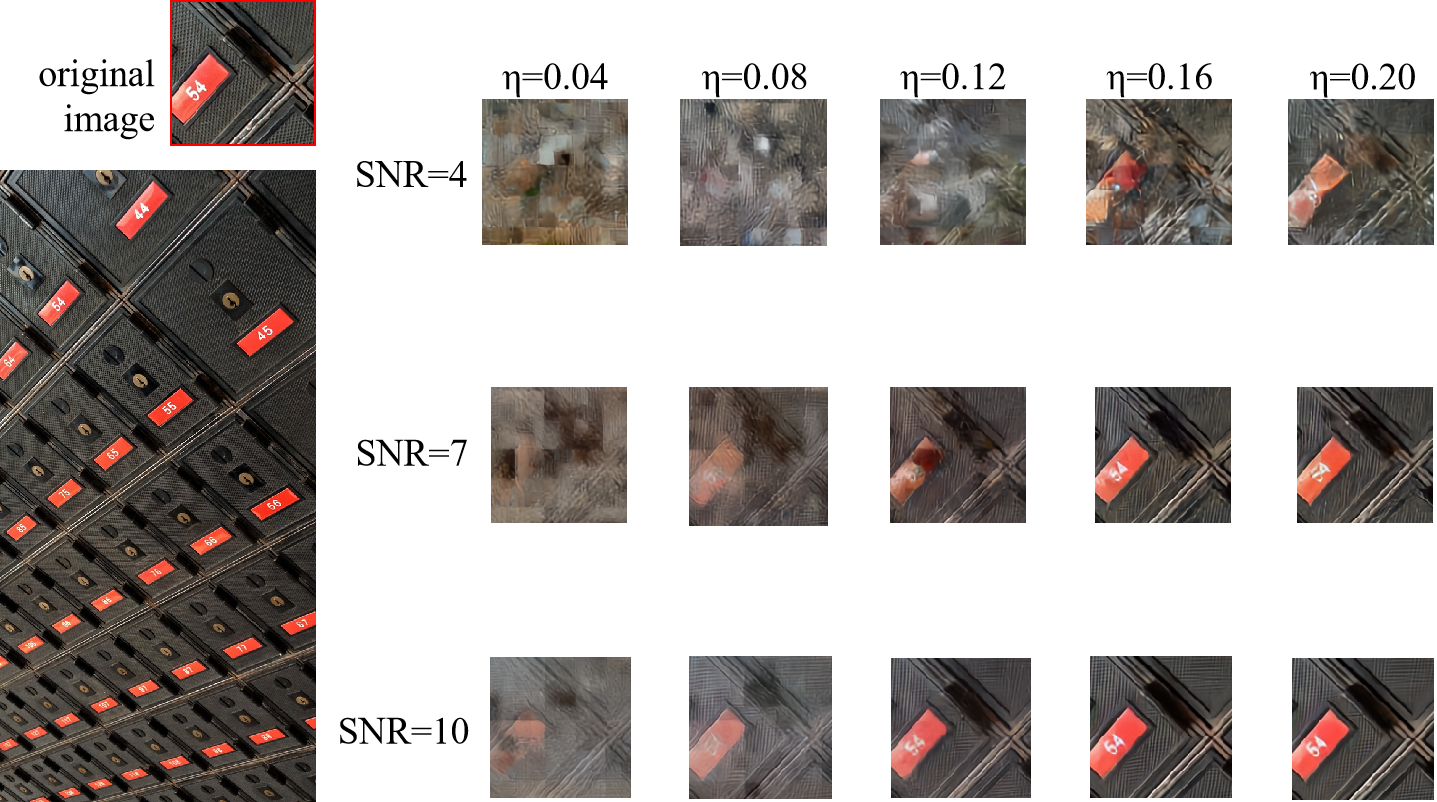}
        \caption{Visualization images with QPSK modulation scheme}
        \label{fig:qpsk}
    \end{subfigure}
    
    \begin{subfigure}{\linewidth}
        \centering
        \includegraphics[width=0.8\linewidth]{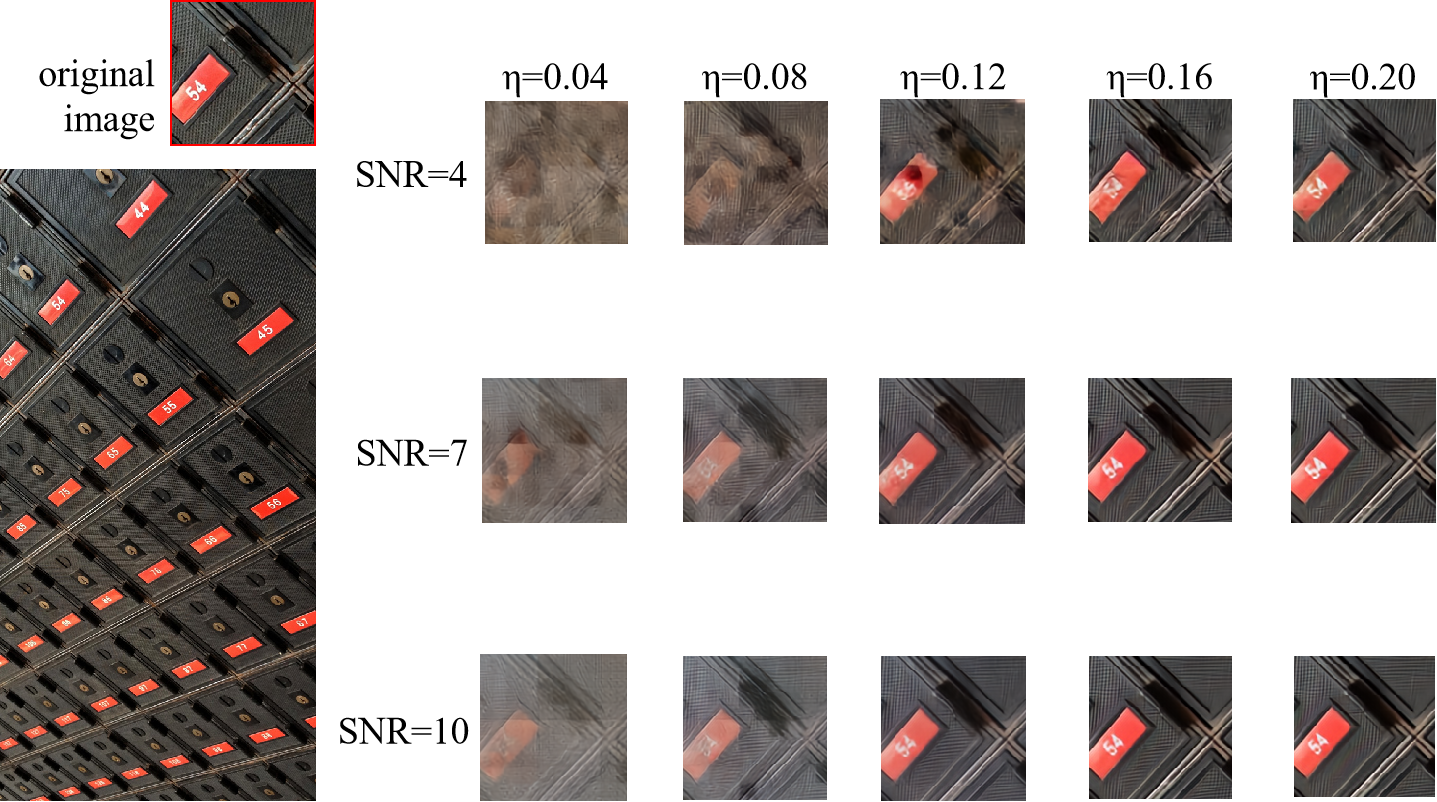}
        \caption{Visualization images with BPSK modulation scheme}
        \label{fig:bpsk}
    \end{subfigure}

    \caption{Visualization images with different $\eta$ and modulation schemes}
    \label{fig:psnr-eta-visual}
\end{figure}

The simulation considers the relationship between the PSNR of reconstructed images and channel SNR for SREC over an AWGN channel under different modulation schemes (16QAM, QPSK, BPSK) and different scaling factors $\eta$ as shown in Figure \ref{fig:psnr-eta-curves} and \ref{fig:psnr-eta-visual}. $\eta$ is used to convert the entropy of semantic features into channel bandwidth cost: a larger $\eta$ indicates a smaller compression ratio, where patches with high entropy are allocated more bandwidth; a smaller $\eta$ indicates a higher compression ratio, where the bandwidth allocated to all patches has smaller differences, but at the cost of sacrificing the reconstruction quality of high-entropy regions. 

From the perspective of modulation schemes, different modulation schemes determine the noise resistance and efficiency during channel transmission. Low-order modulation has strong noise resistance, allowing the potential of compression coding strategies to be fully exploited, resulting in higher image reconstruction quality. In contrast, the high bit error rate of high-order modulation weakens the gain from transmitting more bits. As shown in Figure \ref{eta-16qam},  the PSNR curve is significantly lower across the entire range with a flatter growth trend under 16QAM modulation. Due to the high bit error rate and weak noise resistance of 16QAM, its performance improvement is limited when $\eta$ increases, because many of the additionally transmitted feature bits are corrupted by noise in the channel. Channel SNR is the main factor limiting its performance, and a higher SNR is required to achieve better reconstruction results than BPSK and QPSK. As shown in Figure \ref{eta-bpsk}, the PSNR curve has the maximum slope when $\eta$ is around 0.12 under BPSK modulation, with a significant improvement in reconstruction quality, but the curve tends to saturate after $\eta$ reaches 0.16. Meanwhile, due to BPSK’s excellent noise resistance, the improvement in SNR has almost no impact on PSNR under good channel conditions; reconstruction distortion is mainly limited by the loss of compressed information rather than channel noise. Figure \ref{eta-qpsk} shows that the growth trend of QPSK’s PSNR with $\eta$ is similar to that of BPSK: it has the maximum slope when $\eta$ is around 0.12 and tends to saturate after $\eta$ reaches 0.16. However, QPSK’s PSNR increases significantly with SNR, as the more bits carried per symbol bring more details needed for reconstruction.

\section{Conclusion}
This paper proposes SREC to secure SemCom. This method innovatively integrates an encryption method based on modulo-256 operations, which aims to effectively disrupt transmitted semantic features without introducing the complexity of standard encryption algorithms, preventing unauthorized recovery by eavesdroppers. Meanwhile, a super-resolution image reconstruction module is introduced on the receiver to compensate for the loss of image details caused by channel noise and encryption and decryption processes, thereby improving visual quality. We conduct experiments on the Urban100 dataset to demonstrate the effectiveness of SREC.

\bibliographystyle{IEEEtran}
\bibliography{IEEEabrv,bib}

% Generated by IEEEtran.bst, version: 1.14 (2015/08/26)
\begin{thebibliography}{10}
\providecommand{\url}[1]{#1}
\csname url@samestyle\endcsname
\providecommand{\newblock}{\relax}
\providecommand{\bibinfo}[2]{#2}
\providecommand{\BIBentrySTDinterwordspacing}{\spaceskip=0pt\relax}
\providecommand{\BIBentryALTinterwordstretchfactor}{4}
\providecommand{\BIBentryALTinterwordspacing}{\spaceskip=\fontdimen2\font plus
\BIBentryALTinterwordstretchfactor\fontdimen3\font minus \fontdimen4\font\relax}
\providecommand{\BIBforeignlanguage}[2]{{%
\expandafter\ifx\csname l@#1\endcsname\relax
\typeout{** WARNING: IEEEtran.bst: No hyphenation pattern has been}%
\typeout{** loaded for the language `#1'. Using the pattern for}%
\typeout{** the default language instead.}%
\else
\language=\csname l@#1\endcsname
\fi
#2}}
\providecommand{\BIBdecl}{\relax}
\BIBdecl

\bibitem{trichias20246g}
K.~Trichias, A.~Kaloxylos, and C.~Willcock, ``6g global landscape: A comparative analysis of 6g targets and technological trends,'' in \emph{2024 Joint European Conference on Networks and Communications \& 6G Summit (EuCNC/6G Summit)}, 2024, pp. 1--6.

\bibitem{cao2025importance}
H.~Cao, R.~Meng, X.~Xu, S.~Han, and P.~Zhang, ``Importance-aware robust semantic transmission for leo satellite-ground communication,'' \emph{arXiv preprint arXiv:2508.11457}, 2025.

\bibitem{wang2024intellicise}
W.~Yining, H.~Shujun, X.~Xiaodong, M.~Rui, L.~Haotai, D.~Chen, and Z.~Ping, ``Intellicise model transmission for semantic communication in intelligence-native 6g networks,'' \emph{China Communications}, vol.~21, no.~7, pp. 95--112, 2024.

\bibitem{lu2025important}
H.~Lu, R.~Meng, X.~Xu, Y.~Liu, P.~Zhang, and D.~Niyato, ``Important bit prefix m-ary quadrature amplitude modulation for semantic communications,'' \emph{arXiv preprint arXiv:2508.11351}, 2025.

\bibitem{zhang2025intellicise}
P.~Zhang, W.~Xu, Y.~Liu, X.~Qin, K.~Niu, S.~Cui, G.~Shi, Z.~Qin, X.~Xu, F.~Wang, Y.~Meng, C.~Dong, J.~Dai, Q.~Yang, Y.~Sun, D.~Gao, H.~Gao, S.~Han, and X.~Song, ``semantic communication,'' \emph{IEEE Communications Surveys \& Tutorials}, vol.~27, no.~3, pp. 2051--2084, 2025.

\bibitem{bourtsoulatze2019deep}
E.~Bourtsoulatze, D.~Burth~Kurka, and D.~Gündüz, ``Deep joint source-channel coding for wireless image transmission,'' \emph{IEEE Transactions on Cognitive Communications and Networking}, vol.~5, no.~3, pp. 567--579, 2019.

\bibitem{wu2025lotterycodec}
H.~Wu, G.~Chen, P.~L. Dragotti, and D.~G{\"u}nd{\"u}z, ``Lotterycodec: Searching the implicit representation in a random network for low-complexity image compression,'' \emph{arXiv preprint arXiv:2507.01204}, 2025.

\bibitem{do2025security}
Q.~T. Do, D.~Won, T.~S. Do, T.~P. Truong, and S.~Cho, ``Security and privacy challenges in semantic communication networks,'' in \emph{2025 International Conference on Artificial Intelligence in Information and Communication (ICAIIC)}, 2025, pp. 0032--0035.

\bibitem{qin2023securing}
Q.~Qin, Y.~Rong, G.~Nan, S.~Wu, X.~Zhang, Q.~Cui, and X.~Tao, ``Securing semantic communications with physical-layer semantic encryption and obfuscation,'' in \emph{ICC 2023 - IEEE International Conference on Communications}, 2023, pp. 5608--5613.

\bibitem{rong2025semantic}
Y.~Rong, G.~Nan, M.~Zhang, S.~Chen, S.~Wang, X.~Zhang, N.~Ma, S.~Gong, Z.~Yang, Q.~Cui, X.~Tao, and T.~Q.~S. Quek, ``Semantic entropy can simultaneously benefit transmission efficiency and channel security of wireless semantic communications,'' \emph{IEEE Transactions on Information Forensics and Security}, vol.~20, pp. 2067--2082, 2025.

\bibitem{wu2025actions}
H.~Wu, G.~Chen, and D.~G{\"u}nd{\"u}z, ``Actions speak louder than words: Rate-reward trade-off in markov decision processes,'' \emph{arXiv preprint arXiv:2502.03335}, 2025.

\bibitem{tung2023deep}
T.-Y. Tung and D.~Gündüz, ``Deep joint source-channel and encryption coding: Secure semantic communications,'' in \emph{ICC 2023 - IEEE International Conference on Communications}, 2023, pp. 5620--5625.

\bibitem{meng2025secure}
R.~Meng, D.~Fan, H.~Gao, Y.~Yuan, B.~Wang, X.~Xu, M.~Sun, C.~Dong, X.~Tao, P.~Zhang \emph{et~al.}, ``Secure semantic communication with homomorphic encryption,'' \emph{arXiv preprint arXiv:2501.10182}, 2025.

\bibitem{kaewpuang2024cooperative}
R.~Kaewpuang, M.~Xu, W.~Y.~B. Lim, D.~Niyato, H.~Yu, J.~Kang, and X.~Shen, ``Cooperative resource management in quantum key distribution (qkd) networks for semantic communication,'' \emph{IEEE Internet of Things Journal}, vol.~11, no.~3, pp. 4454--4469, 2024.

\bibitem{zhao2022semkey}
R.~Zhao, Q.~Qin, N.~Xu, G.~Nan, Q.~Cui, and X.~Tao, ``Semkey: Boosting secret key generation for ris-assisted semantic communication systems,'' in \emph{2022 IEEE 96th Vehicular Technology Conference (VTC2022-Fall)}, 2022, pp. 1--5.

\bibitem{xu2024covert}
R.~Xu, G.~Li, Z.~Yang, J.~Kang, X.~Zhang, and J.~Li, ``Covert uav data transmission via semantic communication: A drl-driven joint position and power optimization method,'' in \emph{2024 IEEE/CIC International Conference on Communications in China (ICCC)}, 2024, pp. 66--71.

\bibitem{tang2024secure}
S.~Tang, C.~Liu, Q.~Yang, S.~He, and D.~Niyato, ``Secure semantic communication for image transmission in the presence of eavesdroppers,'' in \emph{GLOBECOM 2024 - 2024 IEEE Global Communications Conference}, 2024, pp. 2172--2177.

\bibitem{yang2020a}
Z.~Yang, P.~Shi, and D.~Pan, ``A survey of super-resolution based on deep learning,'' in \emph{2020 International Conference on Culture-oriented Science \& Technology (ICCST)}, 2020, pp. 514--518.

\bibitem{wu2024seesr}
R.~Wu, T.~Yang, L.~Sun, Z.~Zhang, S.~Li, and L.~Zhang, ``Seesr: Towards semantics-aware real-world image super-resolution,'' in \emph{2024 IEEE/CVF Conference on Computer Vision and Pattern Recognition (CVPR)}, 2024, pp. 25\,456--25\,467.

\bibitem{huang2015single}
J.-B. Huang, A.~Singh, and N.~Ahuja, ``Single image super-resolution from transformed self-exemplars,'' in \emph{Proceedings of the IEEE conference on computer vision and pattern recognition}, 2015, pp. 5197--5206.

\bibitem{wang2023a}
C.~Wang, T.~Zhang, H.~Chen, Q.~Huang, J.~Ni, and X.~Zhang, ``A novel encryption-then-lossy-compression scheme of color images using customized residual dense spatial network,'' \emph{IEEE Transactions on Multimedia}, vol.~25, pp. 4026--4040, 2023.

\bibitem{zhang2012scalable}
X.~Zhang, G.~Feng, Y.~Ren, and Z.~Qian, ``Scalable coding of encrypted images,'' \emph{IEEE transactions on image processing}, vol.~21, no.~6, pp. 3108--3114, 2012.

\bibitem{zhang2018residual}
Y.~Zhang, Y.~Tian, Y.~Kong, B.~Zhong, and Y.~Fu, ``Residual dense network for image super-resolution,'' in \emph{Proceedings of the IEEE conference on computer vision and pattern recognition}, 2018, pp. 2472--2481.

\bibitem{agustsson2017ntire}
E.~Agustsson and R.~Timofte, ``Ntire 2017 challenge on single image super-resolution: Dataset and study,'' in \emph{2017 IEEE Conference on Computer Vision and Pattern Recognition Workshops (CVPRW)}, 2017, pp. 1122--1131.

\bibitem{dai2022nonlinear}
J.~Dai, S.~Wang, K.~Tan, Z.~Si, X.~Qin, K.~Niu, and P.~Zhang, ``Nonlinear transform source-channel coding for semantic communications,'' \emph{IEEE Journal on Selected Areas in Communications}, vol.~40, no.~8, pp. 2300--2316, 2022.

\end{thebibliography}

\end{document}